\documentclass[aps, prd, showpacs, showkeys, preprint]{revtex4}
\usepackage{epsfig}
\begin{document}
\begin{titlepage}
~\vskip1cm
\title{On the infrared freezing of perturbative QCD \\ 
in the Minkowskian region}
\author{Irinel Caprini\footnote{caprini@theory.nipne.ro}}
\affiliation{National Institute of Physics and Nuclear Engineering,Bucharest POB MG-6, R-077125 Romania}
\author{Jan Fischer\footnote{fischer@fzu.cz}}
\affiliation{Institute of Physics, Academy of Sciences of the Czech Republic,CZ-182 21  Prague 8, Czech Republic}

\begin{abstract} The infrared freezing of observables is known to hold  at fixed orders of perturbative 
QCD  if the Minkowskian quantities are  defined through the analytic continuation from the Euclidean region.
 In  a recent paper \cite{HoMa} it is claimed that infrared freezing  can be proved also for Borel resummed
 all-orders quantities in perturbative QCD. In the present paper we obtain the Minkowskian quantities by
 the analytic continuation of the all-orders Euclidean amplitudes expressed in terms of the inverse  Mellin
 transform of the corresponding Borel functions  \cite{CaNe}. Our  result shows that if the principle of
 analytic continuation  is preserved  in Borel-type resummations, the Minkowskian quantities exhibit a
 divergent  increase in the infrared regime, which contradicts the claim made in  \cite{HoMa}. We
 discuss the arguments given in \cite{HoMa} and show that the   special redefinition of Borel summation 
at low energies adopted there does  not reproduce the lowest order result obtained by analytic continuation.
\end{abstract}
\maketitle
 \end{titlepage} 
\newpage\section{Introduction}
Since the advent of QCD it was realized that the application of the renormalization-group improved  
perturbation theory is natural in the deep Euclidean region, where the running coupling is small and 
the physical hadronic thresholds are absent. The application of perturbative QCD  for physical 
observables defined as Minkowskian quantities requires the analytic continuation from the spacelike
 to the timelike axis of the complex momentum plane. At high energies, the analytic continuation of the
 strong running coupling $a(-s)$ from  the Euclidean region $s<0$ to the Minkowskian region  $s>0$ can be
 expanded in powers of $1/\ln(s/\Lambda^2)$. So, in the asymptotic region the expansion parameter is the
 same on the spacelike axis and the timelike one. At lower  energies, however,  one must take into account
 the finite terms appearing from the analytic continuation of $\ln (-s/\Lambda^2)\to \ln (s/\Lambda^2)- i\pi$.
 The problem was investigated  in the early 80' by several authors \cite{PeRo}, \cite{Rady}, who tried
 to identify the most natural parameter for the perturbative QCD expansions of timelike observables.  
In  \cite{PeRo} the authors compare the expansion parameters $a(s)$,$|a(-s)|$ and ${\rm Re} a(-s)$ for
 $s>0$ and notice that $|a(-s)|$ seems suitable since it remains finite in the Landau region $s < \Lambda^2$.
 However, the choice of the modulus $|a(-s)|$ as expansion parameter does not absorb all the $\pi^2$ factors
  which arise  from the analytic continuation, as shown  by Radyushkin  \cite{Rady}, who derived explicit
 formulae for the timelike observables to every finite order of perturbative series. The analytic 
continuation was subsequently applied in the perturbative calculation of Minkowskian  quantities 
\cite{GoKaLa} and in phenomenological analyses of inclusive observables like the rates of the processes 
$e^+e^-\to {\rm hadrons}$ and  $\tau\to {\rm hadrons}$, using either low orders of perturbation theory 
 \cite{Pivo}- \cite{LeDiPi} or resummations based on the Borel method \cite{AlNaRi}-\cite{MatNeu}. 
 
 While for a long time the applications of perturbative QCD  in the
region $0<s<\Lambda^2$ were not considered reliable, the interest in the low energies
increased when it was realized that some Minkowskian quantities, obtained in a consistent
way by analytic continuation, remain finite in the timelike infrared limit $s\to 0$. This property,
 called "infrared freezing", was shown to hold in every finite order of perturbation theory
\cite{ShSo}, \cite{MiSo}, and is actually put on the basis of the so-called "analytic perturbation
theory". In this approach  \cite{ShSo}, \cite{MiSo}, the perturbative expansions of the Minkowskian
observables are defined with a regular effective coupling, and the Euclidean quantities are obtained
thereof by means of dispersion relations known to be valid in QCD \cite{Oehme} under
plausible assumptions.  

One may ask  whether the infrared freezing is only a feature of the finite order QCD expansions or it 
survives beyond finite orders. This  is a  nontrivial question, especially since the QCD perturbative 
series  is known to be divergent. In \cite{HoMa}, using the Borel summation of the QCD perturbative
 series in the leading-$\beta_0$ approximation, the authors conclude that the infrared finite limit of
 the Minkowskian observables is valid  also at all orders in perturbative QCD. Since the perturbative 
series of QCD is ambiguous, it is not impossible, in principle, to implement a desired property by a
 suitable summation prescription. It is however natural to require that the procedure respects the 
principle applied to finite orders, which in the present case is  the analytic continuation. In the 
arguments given in \cite{HoMa} this principle is abandoned at some stage. The reason is that the authors 
use a Borel representation expressed as an infinite series of renormalons in the large-$\beta_0$ 
approximation, which does not display the dependence on the momentum in a transparent way. So the
 question of what is the infrared limit of the Minkowskian quantities when defined in a consistent way
 by analytic continuation from the deep Euclidean region, as is done in the case of fixed orders, remains 
open. In the present paper  we address this question. 

To this end, we choose an alternative representation of the Borel-summed Euclidean quantities,
derived in \cite{CaNe}, which is more convenient for the analytic continuation since it explicitly
displays momentum dependence. A remarkable merit of this approach is that we do not need to
represent the Minkowskian  quantity in terms of any expansion parameter $|a(s)|$, $a(|s|)$ or 
${\rm Re} a(s)$ (as in \cite{PeRo}), assuming only that the quantity admits certain integral 
representations as discussed below in Section III. Note that the same technique was applied in \cite{CaFi}
for the analytic continuation in the coupling plane, leading to results consistent with those obtained in
 \cite{Cvetic}.  
 
 As in \cite{HoMa}, we choose as Euclidean quantity the Adler function in massless QCD and 
as  Minkowskian  quantity the spectral function of the polarization function.
 In the next section we briefly review the analytic continuation from the Euclidean to the
Minkowskian region of fixed-order perturbative expansions  in QCD, stressing upon the fact
that a consistent analytic continuation is free of ambiguities. In Section III we perform the 
similar analytic continuation of the whole  Borel-resummed  Adler function, written in a compact form
in \cite{CaNe}, which displays the energy dependence in an explicit way. In this Section we treat in
detail the one loop coupling.  The situation beyond one-loop is discussed briefly Section IV, where we show that our conclusion about the infrared behaviour of the Minkowskian observables remains valid also in this case. We use the analytic
 expression of the two loop coupling derived  recently in \cite{GGK}, \cite{Magr1},
  working in the assumption, true in the real-world QCD, that the Euclidian  coupling is not causal. 
In Section V we review the Borel
summation  presented in \cite{HoMa} and show that it does not reproduce correctly the lowest
order result obtained by analytic continuation in the infrared limit.

\section{Analytic continuation of fixed-order perturbative expansions}
We consider the Adler function in massless QCD defined as
\begin{equation}\label{Adler} 
{\cal D}(s) =- s {{\rm d}\Pi(s)\over {\rm d}s} - 1\,,
\end{equation}
where $\Pi(s)$ is the correlation function of two vector currents.The function $\Pi(s)$ can be obtained from 
${\cal D}(s)$ by logarithmic integration:
\begin{equation}\label{Pidef}   
\Pi(s) = k -\ln (-s)- \int\limits^s{\rm d}\ln(-s')\,{\cal D}(s') \,,
\end{equation}
where $k$ is a constant and the integration is along a contour in the complex plane which starts at a fixed point and 
ends at $s$, without crossing the singularities of the integrand. This definition is consistent with asymptotic freedom 
and the general properties of the QCD Green functions. Causality and unitarity imply that $\Pi(s)$ and ${\cal D}(s)$ are
 real analytic functions in the complex $s$ plane ({\em i.e.} $\Pi(s^*)=\Pi^*(s)$ and ${\cal D}(s^*)={\cal D}^*(s)$), cut
 along the positive real axis from the threshold for hadron production at $s=4 m_\pi^2$ to infinity. Along the cut, 
 the Minkowskian quantity of interest is related to the spectral function ${\rm Im}\,\Pi(s+i\epsilon)$  by
\begin{equation}\label{R} 
{\cal R}(s) ={1\over \pi}\,{\rm Im}\Pi(s+i\epsilon) -1\,.
\end{equation}
Following \cite{HoMa}, we consider the renormalization group improved truncated expansion of the Adler function in  
perturbative QCD 
\begin{equation}\label{Dseries} 
{\cal D}^{(N)}(s) = a(-s) + \sum\limits_{n\ge 1}^N d_n\, a^{n+1}(-s)\,,
\end{equation}
with the one-loop coupling defined as 
\begin{equation}\label{as1loop}   
a(s) = \frac{\alpha_s(s)}{\pi}= \frac{1}{\beta_0\ln(s/\Lambda^2)} \,,
\end{equation}
 where  $\Lambda$ is the QCD scale parameter and $\beta_0=(11 N_c - 2 n_f)/12$ is the first coefficient
 of the $\beta$ function (we follow in general the notations in \cite{HoMa}, except for using $\beta_0=b/2$ instead of $b$).
 In our analysis we shall assume that  $\beta_0$ is positive, which means that infrared freezing does not hold for
 the Euclidian quantities like $D(s)$ for $s<0$. The first coefficients in (\ref{Dseries}), $d_n$, $n\le 3$,
 were calculated in \cite{GoKaLa}. 
 
 Using (\ref{Pidef}) we obtain the polarization amplitude as:
\begin{equation}\label{Pi}    
\Pi^{(N)}(s) = k - \ln(-s)    - \frac{\ln\ln(-s/\Lambda^2)}{\beta_0} +
 \sum\limits_{n\ge 1}^N d_n  \left(\frac{1}{\beta_0}\right)^{n+1} \frac{1}{n \ln^n(-s/\Lambda^2)}\, .
\end{equation}
We recall that the above expressions are derived in the deep Euclidean region $s <- \Lambda^2$ or,
 more generally, for complex values of $s$, with $|s|> \Lambda^2$. In this region the expressions are consistent with 
the general properties derived from field theory, which require that $\Pi(s)$ and ${\cal D}(s)$ must be real for $s<0$.
 The analytic continuations of (\ref{Dseries}) and (\ref{Pi}) to low values of $|s|$ contain however unphysical singularities
 on the  space-like axis, which are absent from the exact amplitudes: ${\cal D}^{(N)}(s)$ has a Landau pole at $s=-\Lambda^2$,
 and $\Pi^{(N)}(s)$  has a Landau cut along the interval  $-\Lambda^2<s<0$. Here we are interested in the imaginary part of 
$\Pi^{(N)}(s)$  on the upper edge of the timelike axis $s>0$. Using (\ref{Pi}), the spectral function (\ref{R}) is obtained, 
at finite orders, as 
\begin{equation}\label{Ranalyt}    
{\cal R}^{(N)}(s) = A_1(s) + \sum\limits_{n\ge 1}^N d_n A_{n+1}(s) \,,
\end{equation} 
where \cite{Rady}, \cite{ShSo}, \cite{HoMa}
\begin{eqnarray}\label{An}   
A_1(s)&=&   \frac{1}{\pi\beta_0}\left[ \arctan(\pi\beta_0 a(s))+ \pi\theta(\Lambda^2-s)\right]
\nonumber\\ A_n(s)&=& \frac{1}{\pi\beta_0}\, \frac{a^{n-1}(s)}{n-1}\, {\rm Im}\left[(1 - i \pi\beta_0 a(s))^{1-n}\right]\,, 
\quad n>1\,,
\end{eqnarray}
with $a(s)$  defined in (\ref{as1loop}). We note that in the first Eq. (\ref{An})
 $\arctan$ denotes the standard function defined in the interval $(-\pi/2,  +\pi/2)$, with $\arctan(0)=0$, and
 the term $\theta(\Lambda^2-s)$ accounts for the fact that  the real part of $\ln (s/\Lambda^2)$ becomes negative 
when $s<\Lambda^2$. Indeed, writing 
 \begin{equation}\label{ln} 
 \ln (-s/\Lambda^2) =\ln (s/\Lambda^2) - i\pi = \sqrt{\ln^2(s/\Lambda^2)+\pi^2}\, \exp[i \phi] 
  \end{equation}
  for $s$ positive and above the cut, one can see that the phase $\phi$ is continuous at $s=\Lambda^2$, where 
it passes to the second quadrant (as shown in \cite{MiSo}, the first Eq. (\ref{An})  may be written also as 
$A_1(s)= 1/\pi\beta_0 \arccos[L/\sqrt{L^2+\pi^2} ]$, where $L=\ln s/\Lambda^2 = 1/(\beta_{0} a(s))$). 

We mention that in some papers the Minkowskian quantity 
${\cal R}(s)$ for $s>\Lambda^2$  is defined in terms of the Adler function through an integral
along an open contour which ends at $s\pm i\epsilon$. Usually, this contour is chosen as the 
circle of radius  $|s|$ centered at the origin, since in this case the integrals of the
finite order expansions can be done analytically \cite{Rady}, \cite{HoMa}. While this procedure is 
suited for $s$ much larger than $\Lambda^2$, for points close to $\Lambda^2$ the result is sensitive
to small deformations of the integration contour,which may or may not include the Landau pole. In 
\cite{HoMa} this ambiguity is solved by an ad-hoc choice  of the branch of the  arctan function which appears after
 integration, so as to lead to infrared freezing for finite order expansions. We stress that the procedure of calculating 
the discontinuity of the polarization function applied in the above Eqs. (\ref{R})-(\ref{Pi}) is free of 
such ambiguities.

As was mentioned in the Introduction, in applications at large  energies one expands the functions $A_n(s)$ in powers of 
the small coupling $a(s)$ defined in (\ref{as1loop}). This gives for ${\cal R}(s)$ \cite{GoKaLa} 
\begin{equation}\label{Anapprox} 	
{\cal R}(s) \sim a(s) + d_{1} a^{2}(s) +  \left(d_{3} -d_1 \frac{\pi^2 \beta_0^2}{3} 
\right) a^3(s) \dots\,. 
\end{equation} 
The approximate expansion of ${\cal R}$ thus obtained can in no way be used at low energies, since 
$a(s)$ becomes infinite at $s= \Lambda^2$. On  the other hand, as Eqs. (\ref{Ranalyt}) and (\ref{An})
imply, the ${\cal R}^{(N)}(s)$ are regular for all $s$,  including  $s= \Lambda^2$,  and have a finite,
 universal infrared limit  \begin{equation}\label{IF}{\cal R}^{(N)}(0)=\frac{1}{\beta_0}\,,\end{equation} 
for $N$ any positive integer.
\section{Analytic continuation of the Borel summed amplitude}
The perturbation expansion (\ref{Dseries}) of ${\cal D}(s)$ in powers of the renormalized coupling
  $a(-s)$ is known to be neither convergent nor Borel summable. We consider the Borel transform
 $B_D(u)$ defined in the standard way in terms of the  perturbative coefficients $d_n$ of ${\cal D}$:
\begin{equation}\label{B} 
B_D(u)=\sum\limits_{n=0}^\infty {d_n\over n!}\, \left({u\over \beta_0}\right)^n\,,
\quad d_0=1\,. 
\end{equation} 
From the $n!$ large order growth of $d_n$ it is known that $B_D(u)$ has singularities 
(ultraviolet and infrared renormalons) on the real axis of the $u$-plane \cite{Beneke}. For 
the Adler function, the ultraviolet renormalons are placed along the range $u\leq u_1$, $u_1=-1$
and the infrared renormalons along $u \geq u_2$,  $u_2= 2$. Due to the infrared renormalons, the 
usual Borel-Laplace integral is not well-defined  and requires an integration prescription. Defining
\begin{equation}\label{pm}   
{\cal D}^{(\pm)}(s)={1\over \beta_0}\,
\int\limits_{{\cal C}_\pm}\!{\rm e}^{-u/(\beta_0 a(-s))} \, B_D(u)\,{\rm d}u\,= 
{1\over \beta_0}\,\lim\limits_{\epsilon\to 0}\,\int\limits_{0\pm i\epsilon}^{\infty\pm i\epsilon}\!
{\rm e}^{-u/(\beta_0 a(-s))} \, B_D(u)\,{\rm d}u\,,
\end{equation}
 one can adopt as prescription, for each value of $a(-s)$ with ${\rm Re}\,
  a(-s)>0$, either ${\cal D}^{(+)}(s)$ or ${\cal D}^{(-)}(s)$,
 or a linear  combination of them, with coefficients $\xi$ and $1-\xi$ such as correctly to 
reproduce the known perturbative (asymptotic) expansion  (\ref{Dseries}) of ${\cal D}(s)$ 
 (obtained by truncating the Taylor expansion (\ref{B}) at a finite order $N$).  Note that all these
 (and many other) integration prescriptions have, according  to a theorem by Watson \cite{Wat}, 
the same asymptotic series in powers of  $a(-s)$ and therefore possess the same, original perturbative 
expansion.

Once a prescription is adopted, one has a well-defined  function, different prescriptions 
yielding different functions with different properties. In the present work we use, as in \cite{HoMa}, 
the principal value ($PV$) prescription
\begin{equation}\label{pv}
{\cal D}(s)={1\over 2} [{\cal D}^{(+)}(s)+ {\cal D}^{(-)}(s)]\,.
\end{equation}
As shown in \cite{CaNe},
 this prescription gives real values along the space-like axis outside the Landau region, which is 
consistent with the general analyticity requirements imposed by causality and unitarity. Moreover, 
we work in the $V$-scheme, where all the exponential dependence in the Laplace integrals (\ref{pm}) is 
absorbed in the running coupling, and denote by $\Lambda_V^2$  the corresponding QCD scale parameter. 

Our purpose is to obtain the Minkowskian qauantity
${\cal R}$ by analytically continuing the Euclidean Borel-summed Adler function
(\ref{pv}). To this end it is convenient to use a representation of the Borel function
$B_D$ in terms of its inverse Mellin transform $\widehat w_D$  defined as \cite{Matt1}
\begin{equation}\label{wD}   
\widehat w_D(\tau) = \frac{1}{2\pi i} 
  \int\limits_{u_0-i\infty}^{u_0+i\infty}\!{\rm d}u\,   B_D(u)\,\tau^{u-1} \,. 
  \end{equation}
The inverse relation 
\begin{equation}\label{wninv}   
B_D(u) = \int\limits_0^\infty\!{\rm d}\tau\, 
  \widehat w_D(\tau)\,\tau^{-u} \,, 
  \end{equation} 
  defines the function $B_D(u)$ in a strip parallel to the
imaginary axis with $u_1<\mbox{Re}\,u<u_2$.  The relations (\ref{wD}) and 
(\ref{wninv}) are valid if the following $L^2$ condition holds \cite{Tit}:\begin{equation}\label{L2B}
   \frac{1}{2\pi i} \int\limits_{u_0-i\infty}^{u_0+i\infty}\!   {\rm d}u\,|B_D(u)|^2 < \infty \,,
\end{equation} 
where $u_0\in[u_1,u_2]$.  The function $\widehat w_D(\tau)$ was calculated in \cite{Matt1}
 in the large-$\beta_0$ approximation \cite{Bene}, \cite{Broa}, where it has different analytic 
expressions,  which we denote by $\widehat w_D^{(<)}$ and $\widehat w_D^{(>)}$, depending on whether 
$\tau$ is less or greater than 1, respectively:
\begin{eqnarray}\label{wDdif}  
\widehat w_D(\tau) &=& \widehat w_D^{(<)}(\tau),\quad 0< \tau<1 \nonumber\\ 
 \widehat w_D(\tau) &=&  \widehat w_D^{(>)}(\tau),\quad \tau>1 \,.
\end{eqnarray}
 As discussed in \cite{CaNe}, one expects the inverse Mellin transform $ \widehat w_D$  
 to have different expressions for $\tau <1$ and $\tau >1$  in general, also beyond the leading
 $\beta_0$-approximation. Indeed,  $\widehat w_D^{(<)}$ is given by a sum over the residua  of 
the infrared renormalons, while $\widehat w_D^{(>)}$ is calculated in terms of the residua of the 
ultraviolet renormalons, and there are no reasons to expect these two contributions  to be equal. 
In the large-$\beta_0$ approximation, the expressions of the functions $\widehat w_D^{(<)}$ and 
$\widehat w_D^{(>)}$ are \cite{Matt1}:
\begin{eqnarray}\label{wDfun}   
\widehat w_D^{(<)}(\tau) &=&
 \frac{8}{3} \left\{ \tau\left(    \frac74 - \ln\tau \right) + (1+\tau)\Big[ L_2(-\tau) + \ln\tau  
  \ln(1+\tau) \Big] \right\} \,,\\   \widehat w_D^{(>)}(\tau) &=& \frac{8}{3} \left\{ 1 + \ln\tau  
  + \left( \frac34 + \frac12 \ln\tau \right) \frac{1}{\tau}    + (1+\tau)\Big[ L_2(-\tau^{-1}) -
 \ln\tau \ln(1+\tau^{-1}) \Big]    \right\} \,,\nonumber
 \end{eqnarray}
where $L_2(x)=-\int_0^x {{\rm d}t\over t}\ln(1-t)$ is the Euler dilogarithm.
\begin{figure}\unitlength=1mm
\begin{picture}(160, 80)\put(14, -0){\psfig{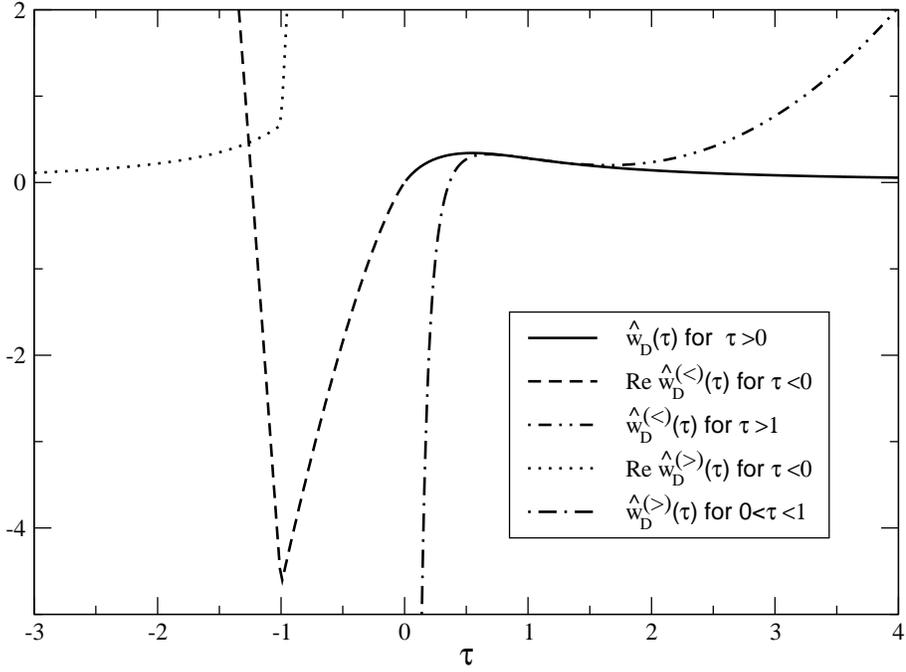}}
\end{picture}
\caption{The function  $\widehat w_D (\tau)$ defined by (\ref{wDdif}) in the large $\beta_0$ 
approximation 
(i.e., with $\widehat w_D^{(<)}(\tau)$ and  $\widehat w^{(>)}_D (\tau)$ given by (\ref{wDfun})) is 
represented by a solid line. We separately display also the function 
$\widehat w_D^{(<)}(\tau)$ for $\tau > 1$ (dash-dot-dotted line) and  for $\tau <0$ (dashed line),
 and the function $\widehat w_D^{(>)}(\tau)$ for $\tau < 1$ 
 (dash-dotted line and dotted line).  
For negative values of $\tau$ the curves represent the real parts of the 
corresponding functions. The dashed line is unbounded for $\tau$ tending to
 $-\infty$, while the dash-dot-dotted line grows unboundedly for $\tau$ increasing. The  dotted and 
dash-dotted lines are unbounded for $\tau$
 approaching zero from either side.}\label{fig:wD}
\end{figure}

As noticed in \cite{Matt1}, the function $\widehat w_D(\tau)$ defined in (\ref{wDdif}) is continuous  
together with its first three derivatives, and satifies the normalization condition:
\begin{equation}\label{norma} 
\int\limits_0^\infty \widehat w_D(\tau)\, {\rm d}\tau =1\,. 
\end{equation}
On the other hand,
  Eqs. (\ref{wDfun}) define two independent functions,  $\widehat w_D^{(<)}$ and  $\widehat w_D^{(>)}$,
 which are analytic in the whole $\tau$ complex plane except for logarithmic branch-points. These functions 
 are  not bounded everywhere:   the function  $\widehat w_D^{(<)}(\tau)$ is unbounded for $\tau>1$ 
(growing at infinity like $\tau \ln^2\tau$), while $\widehat w_D^{(>)}(\tau)$ grows like  $\ln\tau/\tau$ for
 $\tau\to 0$. This behavior is seen in Figure 1, where we represent the function $\widehat w_D (\tau)$ defined by 
(\ref{wDdif}), together with $\widehat w_D^{(<)}(\tau)$ for $\tau >1$ and  $\widehat w_D^{(>)}(\tau)$  for $\tau <1$. 
The same figure shows also the real  parts of the functions $\widehat w_D^{(<)}(\tau)$ and  $\widehat w_D^{(>)} (\tau)$
 for $\tau<0$, where they become complex.  As we will see below, the  growth of ${\rm Re}\, \widehat w_D^{(<)}(\tau)$ for 
$\tau\to-\infty$  will have important consequences for the problem investigated in the present work. 

Following the technique described in detail in \cite{CaNe}, we shall express the function
 ${\cal D}(s)$ for complex values of $s$, with ${\rm Re}a(-s) >0$, in terms of the inverse
 Mellin transform $\widehat w_D$. Using Eqs. (\ref{pm}) and (\ref{pv}) as starting points,
we rotate the integration contours ${\cal C_\pm}$ in the complex $u$-plane up to a line parallel to 
the imaginary axis where the representation (\ref{wninv}) of $B_D$ holds and can be inserted into the Borel integral. 
If the integrals are convergent, we can reverse the order of integration upon $u$ and $\tau$, and perform first the integral
 upon the variable $u$, which can be done exactly. As explained in \cite{CaNe},  when $s$ is in the upper half of the complex 
plane, the contour ${\cal C_+}$ can be rotated towards the positive imaginary axis in the $u$-plane since the integrals remain 
convergent, while for the integral along the contour ${\cal C_-}$ it is necessary to first cross the real positive axis of 
the $u$-plane, picking up contribution of the residua of the corresponding  singularities, 
{\em i.e.} the infrared renormalons.
 When $s$ is in the lower half of the complex plane, convergence is achieved if the contours are be rotated towards the 
negative imaginary axis in the $u$-plane, and the roles of the contours ${\cal C_+}$ and ${\cal C_-}$ are reversed. This  
gives different expressions for ${\cal D}(s)$ in the upper/lower semiplanes of the $s$ plane:
\begin{eqnarray}\label{Duplow}
   {\cal D}(s) =  \frac{1}{\beta_0} \int\limits_0^\infty\!   {\rm d}\tau\,\frac{\widehat w_D(\tau)}{\ln(-\tau s/\Lambda_V^2)} 
   -  \frac{i\pi}{\beta_0} \left( -\frac{\Lambda_V^2}{s} \right)   \widehat w_D^{(<)}(-\Lambda_V^2/s) \,,\quad {\rm Im}s>0\,
\nonumber\\
{\cal D}(s) =  \frac{1}{\beta_0} \int\limits_0^\infty\!   {\rm d}\tau\,\frac{\widehat w_D(\tau)}{\ln(-\tau s/\Lambda_V^2)}
    + \frac{i\pi}{\beta_0} \left( -\frac{\Lambda_V^2}{s} \right)   \widehat w_D^{(<)}(-\Lambda_V^2/s) \,,\quad {\rm Im}s<0,
\end{eqnarray}
where the first terms are given by the integration with respect to $u$, and the last terms are produced by the 
residua of the infrared renormalons picked up by crossing of the positive axis. We recall that the expressions 
 (\ref{Duplow})  were obtained by using the Principal Value prescription (\ref{pv}). 
 
The corresponding expression for 
the polarization function $\Pi(s)$ can be obtained by inserting the above result into the definition (\ref{Pidef}). 
This gives
\begin{equation}\label{Piuplo}   
\Pi(s) = k - \ln\left( \frac{-s}{\Lambda_V^2} \right)   - \frac{1}{\beta_0}
 \int\limits_0^\infty\!{\rm d}\tau\,   \widehat w_D(\tau) \ln\ln\left( -\frac{\tau s}{\Lambda_V^2} 
  \right) \pm \frac{i\pi}{\beta_0}\, \int\limits^s {\rm d} \ln (-s)\left(- \frac{\Lambda_V^2}{s}\right)
\widehat w_{D}^{(<)}(-\Lambda_V^2/s) \,, 
\end{equation} 
with the contour in the last integral  specified below 
Eq. (\ref{Pidef}) and the $+/-$  sign corresponding to ${\rm Im}s>0/ {\rm Im}s<0$, respectively. 
The Borel-summed expression   (\ref{Piuplo}), obtained for $|s|>\Lambda_V^2$, can be analytically continued
 into the whole complex plane. We note that $\Pi(s)$ is holomorphic for complex values of $s$ and satisfies the
 reality condition $\Pi(s^*)=\Pi^*(s)$. On the real axis, this function can have singularities manifested as
 discontinuities of the imaginary part. The unphysical singularities in the spacelike region $-\Lambda_V^2<s<0$ 
were discussed in detail in \cite{CaNe}. Here we are interested in the spectral function for $s>0$. 
A straightforward calculation gives:
\begin{eqnarray}\label{imagpi}   
\mbox{Im}\,\Pi(s+i\epsilon) &=& \pi + \frac{1}{\beta_0} 
   \int\limits_0^\infty\!{\rm d}\tau\,\widehat w_D(\tau)    \arctan\left(\frac{\pi}{\ln(\tau s/\Lambda_V^2)} \right)  
     + \frac{\pi}{\beta_0}\,    \int\limits_0^{\Lambda_V^2/s}\!{\rm d}\tau\,\widehat w_D(\tau) \nonumber\\   
 + && \frac{\pi}{\beta_0}\,\mbox{Re} \int\limits_{-\Lambda_V^2/s}^0\!{\rm d}\tau\,   
 \widehat w_D^{(<)}(\tau)\,. 
 \end{eqnarray}
We note that the last term in the first line was obtained by means of the relation
\begin{equation} {\rm Im} \left[\ln\ln\left(-\frac{\tau s}{\Lambda_V^2}\right)\right]=\arctan\left(\frac{\pi}
{\ln(\tau s/\Lambda_V^2)}\right) + \pi\,\theta (\Lambda_V^2- \tau s)\,, \end{equation}
already applied (for $\tau=1$) in deriving the first relation (\ref{An}), and the last term in (\ref{imagpi}) 
is produced to the last term in (\ref{Piuplo}). Using (\ref{R}) we write also ${\cal R}$ as
\begin{equation}\label{Rsummed}   
{\cal R}(s)= \frac{1}{\pi \beta_0}    \int\limits_0^\infty\!{\rm d}\tau\,\widehat w_D(\tau)
    \arctan\left(\frac{\pi}{\ln(\tau s/\Lambda_V^2)} \right)       + \frac{1}{\beta_0}    \int\limits_0^{\Lambda_V^2/s}
\!{\rm d}\tau\,\widehat w_D(\tau)    + \frac{1}{\beta_0}\mbox{Re} \int\limits_{-\Lambda_V^2/s}^0\!{\rm d}\tau\, 
 \widehat w_D^{(<)}(\tau)\,. 
 \end{equation} 
 It is easy to check that this expression is continuous for all $s>0$, 
including the point  $s=\Lambda_V^2$. We consider now the limit of this expression for $s\to 0$, i.e. $\Lambda_V^2/s\to\infty$.
 Since the first integral tends to zero (recall the comment below Eq. (\ref{An}) about arctan) and  
$\widehat w_D(\tau)$ satisfies the normalisation relation (\ref{norma}), we obtain: 
\begin{equation}
\label{Rinfra} 
{\cal R}(0)= \frac{1}{\beta_0} + \lim\limits_{s\to 0} \mbox{Re} \int\limits_{-\Lambda_V^2/s}^0\!{\rm d}\tau\,  
  \widehat w_D^{(<)}(\tau)\,. 
  \end{equation} 
  The first term coincides with the result (\ref{IF}), but we have now an
 additional term which involves the values of $\widehat w_D^{(<)}(\tau)$ for negative $\tau$.  Moreover, for small 
values of $s$, the integral involves arguments $\tau\to-\infty$,  where  ${\rm Re}\,\widehat w_D^{(<)}(\tau)$  
 is unbounded (see Figure 1). Using  (\ref{wDfun}) (which is valid in the large $\beta_0$ limit) it is easy to check 
that the last integral in (\ref{Rinfra}) diverges like  $\ln^2 s /s^2$ for $s\to 0$. This result disproves the
 statement made in \cite{HoMa}: the infrared limit of the Borel-summed Minkowskian observable ${\cal R}(s)$ obtained by
 analytic continuation from the Euclidian region does not reproduce the infrared freezing observed in the finite orders,
 and moreover displays an unphysical divergence. This behavior is crucially determined by the summation of an infinity of
 terms of the series. 
 
 In order to understand the transition from the fixed orders to the resummed quantity, it is useful
 to apply the above formalism to the truncated perturbative expansion,  when the Borel transform $B_D(u)$ defined in (\ref{B})
 reduces to a polynomial $B_D^{(N)}(u)$:
\begin{equation}\label{BN}B_D^{(N)}(u)=1 + d_1 \frac{u}{\beta_0} +  \frac{d_2}{2!}\,
 \frac{u^2}{\beta_0^2}+\ldots  +  \frac{d_N}{N!}\, \frac{u^N}{\beta_0^N}\,. \end{equation} 
In this case the Laplace-Borel
 transform is well defined on the cuts. A straightforward calculation gives 
\begin{equation}\label{Laplace}   
{1\over \beta_0}\,\int\limits_0^\infty\!{\rm e}^{-u/(\beta_0 a(-s))}
 \, B^{(N)}_D(u)\,{\rm d}u\,= \sum\limits_{n=0}^N d_n a^{n+1}(-s) \,,\quad d_0=1, 
 \end{equation}
{\em i.e.} the finite order expansion (\ref{Dseries}). We want to check whether this expansion, along with 
the expansion (\ref{Ranalyt}) of ${\cal R}$, are reproduced in the inverse Mellin formalism. 

Clearly, when $B_D$ 
is a polynomial the condition (\ref{L2B}) is not satisfied, so we expect the function $\widehat w_D$ to be a 
 generalized function (a distribution). In order to calculate it, we consider the alternative distribution function 
$\widehat W_D(\tau)$, introduced in \cite{BBB}: 
\begin{equation}\label{WD}   
\widehat W_D(\tau) = \frac{1}{2\pi i} 
 \int\limits_{u_0-i\infty}^{u_0+i\infty}\!{\rm d}u\,  \frac{B_D(u)}{\sin\pi u}\,\tau^{u-1} \,. 
 \end{equation}
When $B_D(u)$ is a polynomial, the ratio $B_D(u)/\sin\pi u$ satisfies  the condition (\ref{L2B}),
 which ensures the existence of the  function $\widehat W_D(\tau)$. For instance, for $B_D(u)=1$ a 
straightforward calculation gives
\begin{equation}\label{WD1}  \widehat W_D(\tau) =\frac{1}{\pi (1+\tau)}\,.\end{equation}
On the other hand, as shown in \cite{BBB}, \cite{Matt2} (see also Eq. (A.10) of \cite{CaNe}),
 the connection between the Mellin transforms $\widehat W_D(\tau)$ and  $\widehat w_D(\tau)$ is
\begin{equation}\label{rel}   
\widehat W_D(\tau) = \frac{1}{\pi} \int\limits_0^\infty\!{\rm d}x\, 
  \frac{\widehat w_D(x)}{x+\tau} \,, 
  \end{equation}where $\tau$ can take arbitrary values, except for real
 negatives. By comparing (\ref{WD1}) and (\ref{rel}) it follows that, for  $B_D^{(0)}(u)=1$,
\begin{equation}\label{wD2}\widehat w_D^{(0)}(x)=\delta(1-x)\,.\end{equation}
 A straightforward calculation shows that at each finite order the function $\widehat w_D(x)$ is
 represented in terms of the  distribution $\delta(1-x)$ and its derivatives. For instance, the inverse Mellin
 transform of $B_D^{(N)}$ 
defined (\ref{BN}),  for $N=3$, is 
\begin{equation} \label{wD3} 
\widehat w_D^{(3)}(x)=\delta(1-x)-\frac{d_1}{\beta_0}
 \delta'(1-x)+\frac{d_2}{2!\beta^2_0}[ \delta''(1-x) +\delta'(1-x)]-\frac{d_3}{3!\beta^3_0}[ \delta'''(1-x) + 3\delta''(1-x)+
 \delta'(1-x)]\,.
 \end{equation}
 Such a representation is not unique: except for the first two terms which remain the same,
 the higher terms can bewritten  equivalently as the product of the
 $n$-th derivative of $\delta(1-x)$ with a polynomial of degree $n-1$.  For instance, $\widehat w_D^{(3)}(x)$ in (\ref{wD3}) 
can be expressed in the form
\begin{equation}\label{wD4} 
\widehat w_D^{(3)}(x)=\delta(1-x)-\frac{d_1}{\beta_0} \delta'(1-x)+
\frac{d_2}{2!\beta^2_0}[ (x+1)/2  \,\,\,\delta''(1-x)]-\frac{d_3}{3!\beta^3_0}[ (x^2 + 4x + 1)/6 \,\,\, \delta'''(1-x)]\,.
\end{equation} 
It is easy to check that the different expressions (\ref{wD3}) and (\ref{wD4}) give  the same result for 
the quantities of interest ${\cal D}$ and ${\cal R}$. An immediate consequence of these expressions is that, for finite 
orders,  $\widehat w^{(N)<}_D(\tau)=0$. Inserting $\widehat w_D^{(3)}$ into the relation (\ref{Duplow}), which expresses 
the Adler function in terms of the inverse Mellin transform, we obtain by a straightforward calculation
 \begin{equation}\label{Duplow1}   
 {\cal D}^{(3)}(s) =  \frac{1}{\beta_0}   \frac{1}{\ln(- s/\Lambda_V^2)} + d_1 
 \left(\frac{1}{\beta_0}   \frac{1}{\ln(- s/\Lambda_V^2)}\right)^2+ d_2  \left(\frac{1}{\beta_0} 
  \frac{1}{\ln(- s/\Lambda_V^2)}\right)^3+ d_3 \left(\frac{1}{\beta_0}   \frac{1}{\ln(- s/\Lambda_V^2)}\right)^4,
\end{equation} 
which coincides with the first terms in the expansion (\ref{Dseries}).

Let us insert also (\ref{wD3}) into  the resummed expression (\ref{Rsummed}) of ${\cal R}$. It
 is easy to see that the first term $\widehat w_D^{(0)}(\tau)=\delta(1-\tau)$ contributes both to the first and
 the second integrals in  (\ref{Rsummed}), giving the result
\begin{eqnarray}\label{R0}  
{\cal R}^{(0)}(s)&=& \frac{1}{\pi\beta_0}\left[\int_{0}^{\infty} \delta(1-x)
 \arctan\left(\pi/(\ln x + \ln(s/\Lambda^2))\right){\rm d}x +  \pi \int_{0}^{\Lambda^2/s}\delta(1-x){\rm d}x \right]
 \nonumber \\ &=&   \frac{1}{\pi\beta_0}\left[ \arctan(\pi\beta_0 a(s))+ \pi\theta(\Lambda^2-s)\right], 
 \end{eqnarray}
which coincides with the function $A_1(s)$ defined in (\ref{An}) and satisfies
  the property of infrared freezing. The higher terms in (\ref{wD3}) contribute only to the first integral in (\ref{Rsummed}),
 reproducing the terms in expression (\ref{Rsummed}).  For instance, inserting the second term
 of  (\ref{wD3}) in (\ref{Rsummed}) one has
\begin{equation}\label{R1}   
-\frac{d_1}{\pi\beta^2_0}\left[\int_{0}^{\infty} 
\delta'(1-x) \arctan\left(\pi/(\ln x + \ln(s/\Lambda^2))\right){\rm d}x   \right] =  d_1 \frac{a^2}{1 + a^{2} \beta_{0}^{2}
 \pi^{2}} = d_1 A_2(s), 
 \end{equation}
with $A_2(s)$ defined in (\ref{An}). So the formalism of inverse Mellin transform reproduces the finite 
order expansions which are consistent with the property of infrared freezing (\ref{IF}). But the summation 
of the whole series leads to a different result. The discussion in this Section   reveals the difference 
between the finite orders and the summed expression: it resides in the function $\widehat w^{<}_D(\tau)$,
 which is zero at each finite order but is nonvanishing when an infinity of terms are summed and the
 infrared renormalons show up.
\section{Beyond the one-loop coupling}
Up to now we restricted the discussion to the one-loop coupling (\ref{as1loop}). We show now that the same
 conclusion is valid beyond this approximation. If the one loop coupling is not inserted into (\ref{pm}),
 it is easy to see that the two integrals in (\ref{Piuplo}) write in general 
\begin{equation}\label{Piuplo2} 
\Pi(s) \sim   - \frac{1}{\beta_0} \int\limits^s
 {\rm d} \ln(-s)\int\limits_0^\infty\!{\rm d}\tau\,   \frac{\widehat w_D(\tau)}
{\ln\tau +\frac{1}{\beta_0\,a(-s)}} \,    \pm \frac{i\pi}{\beta_0}\, \int\limits^s {\rm d}
 \ln (-s)e^{-\frac{1}{\beta_0 a(-s)}}\widehat w_{D}^{(<)}\left(e^{-\frac{1}{\beta_0 a(-s)}}\right) \,.
\end{equation} 
As shown recently \cite{GGK}, \cite{Magr1} the solution of the two-loop $\beta$-function equation can be written
 analytically in closed form as 
 \begin{equation}\label{as2loop} 
 a(s)=-\frac{1}{c\, [1+ W(z(s))]}\,,
\quad\quad z(s)=-\frac{1}{e}\left(\frac{s}{\Lambda^2}\right)^{-\beta_0/c}\,, 
\end{equation}
where $c=(153 -19 n_f)/24\beta_0$ is the second universal beta-function coefficient  and $W(z)$ is the Lambert
function defined implicitly by $W(z) e^{W(z)}=z$ \cite{Lambert}. 
 We work in the condition   $c>0$, valid in real-world QCD. Then, as shown in  \cite{GGK},
  the physical branch  of the Lambert function in  (\ref{as2loop}) is $W_{-1}$, and  the coupling
 does not freeze in the spacelike region. 

Denoting  $W(z)=W_{-1}(z(-s))$  we obtain  from (\ref{as2loop}):
\begin{eqnarray}\left(\frac{s} 
{\Lambda^2}\right)^{-\beta_0/c} = - W(z) e^{W(z)+1}  \nonumber
 \\\ln \left(\frac{s}{\Lambda^2}\right) = -\frac{c}{\beta_{0}}\left(\ln(-1) + \ln z(s) + 1 \right)\label{chanvar} 
\end{eqnarray} 
and 
\begin{equation}{\rm d} \ln (-s) = -\frac{c}{\beta_{0}}{\rm d}\ln z(-s) 
=-\frac{c}{\beta_{0}}\left({\rm d} 
\ln W + {\rm d} W \right) = -\frac{c}{\beta_{0}}\frac{1 + W}{W}{\rm d} W. \label{subst}
\end{equation}
We now insert the two loop coupling (\ref{as2loop})
 into (\ref{Piuplo2}), and notice that the first integral can be performed by making the change of variable (\ref{subst}).
 Using the relation $W'(z)= W(z)/z(1+W(z))$ (to be obtained from $We^{W}=z$ by differentiating the logarithm)
 we write after a 
straightforward calculation  the first integral in (\ref{Piuplo2}) as
\begin{equation}\label{int1} 
\Pi\sim  - \frac{1}{\beta_0}\int\limits_0^\infty\!{\rm d}\tau\, 
 \widehat w_D(\tau)\left[ \ln\left(1+W(z)-\frac{\beta_0}{c} \ln\tau\right)+\frac{1}{1-\frac{\beta_0}{c}\ln\tau}
\ln \frac{W(z)}{ 1+W(z)-\frac{\beta_0}{c} \ln\tau}\right].
\end{equation}
 In order to evaluate the limit $s\to 0$, we use the asymptotic expansion \cite{Lambert}
\begin{equation}\label{Was} 
W(z)\sim \ln z -\ln \ln z 
\end{equation}
valid for both $z \to 0$ and $z \to \infty$. 
Then we obtain from the first logarithm  in (\ref{int1}) the limit
\begin{equation}\label{int11} 
\lim\limits_{s\to 0} \Pi(s)\sim  - \frac{1}{\beta_0}\int\limits_0^\infty\!{\rm d}\tau\, 
 \widehat w_D(\tau) \,\ln\ln\left(\frac{-s \tau}{\Lambda^2}\right)\,, 
 \end{equation}while the second term in (\ref{int1}) 
vanishes.The expression (\ref{int11}) coincides with the first integral in (\ref{Piuplo}), whose imaginary part, as 
discussed in Eqs. (\ref{imagpi})-(\ref{Rinfra}), has the infrared limit $1/\beta_0$.  This result is consistent with 
the finite order calculations with the two loop coupling in \cite{HoMa}, \cite{Magr2}. 

We turn now to the second integral 
in (\ref{Piuplo2}). It is easy to see that the leading term in the asymptotic behavior (\ref{Was}) gives, for $s\to 0$, 
 an expression identical to the second term in (\ref{Rinfra}). The next to leading term in (\ref{Was}) introduces logarithmic 
corrections which do not change the singular behavior of the integrand for $s\to 0$, {\em i.e.} $\tau\to -\infty$. Thus, 
although the two loop coupling leads to a different behavior of the spectral function at nonvanishing $s$, the singularity
 in the infrared limit is dominated by the one loop coupling.
\section{Comments on the proof of infrared freezing in \cite{HoMa}}
In Ref. \cite{HoMa} the authors discuss the infrared freezing beyond  fixed-order expansions 
 by using a representation  of the Minkowskian quantity  ${\cal R}$ in terms of the Borel transform $B_D$ of
 the Adler function. This  representation is derived by  inserting the Borel representation of the Adler  function  
into the expression (\ref{Pi}) of the polarization function, which  is then used in (\ref{R}) to compute  the quantity 
 ${\cal R}$. Adopting  the Principal Value prescription   (\ref{pv}), a straightforward calculation  gives \cite{AlNaRi}, 
\cite{HoMa}:
\begin{equation}\label{RBD} 
{\cal R}(s)= \frac{1}{\beta_0} {\rm P V} \int\limits_{0}^{\infty} {\rm e}^{-u/(\beta_0 a(s))} 
\,\frac{\sin\pi u}{\pi u}\, B_D(u) {\rm d} u,
 \end{equation}
with the one-loop coupling $a(s)$ defined in (\ref{as1loop}). 

The integrals in (\ref{RBD}) 
converge only for $s>\Lambda^2$, when $a(s)>0$. For small $s$, of interest for the infrared behavior, $a(s)$ is 
 negative and the standard Laplace-Borel integral, along the positive real axis of the $u$-plane, diverges.  
In  \cite{HoMa} the authors notice that for   $a(s)<0$  a convergent integral is obtained  by choosing as integration 
line the negative axis in the $u$-plane, instead of the positive one. Therefore, they define the Borel-summed
  ${\cal R}(s)$ for  $s<\Lambda^2$ as:
 \begin{equation}\label{RBDm} 
 {\cal R}(s)= \frac{1}{\beta_0}{\rm PV} \int\limits_{0}^{-\infty} {\rm e}^{-u/(\beta_0 a(s))} 
\,\frac{\sin\pi u}{\pi u}\, B_D(u) {\rm d} u .
\end{equation}
The PV prescription now regulates the ultraviolet renormalons along the negative real axis, 
while in the standard definition (\ref{RBD}) the prescription regularizes the infrared renormalons. 
The authors of \cite{HoMa} claim that by this redefinition of the Minkowskian quantity for $s<\Lambda^2$, 
one recovers infrared freezing (\ref{IF}) also beyond fixed-order perturbation expansions. 

In principle, 
a redefinition of the Borel integral as in (\ref{RBDm}) is not illegitimate: since we deal with functions
 which are not Borel summable, the choice of the prescription for the Borel integral is to a large extent arbitrary; 
as mentioned in section III., there are many different functions that have the same perturbative, divergent asymptotic
 expansion in powers  of the coupling constant. Such a redefinition should however be motivated physically, not just by 
the fact that the original definition has lost mathematical sense in a region. 

Furthermore, when introducing a redefinition 
of the Borel integral, we must ensure that the fixed order expansions are reproduced when the Borel transform is expanded in 
a Taylor series. In \cite{HoMa} the authors claim this requirement to be fulfilled if  one  expands in powers of $u$ not the
 whole integrand in (\ref{RBD}), but only the Borel transform $B_D(u)$.   However, it is easy to see that 
 the prescription (\ref{RBD1}) fails to reproduce correctly the infrared freezing (\ref{IF}) for finite order expansions. 

Indeed, let us insert in (\ref{RBDm}) the truncated expansion (\ref{BN}) of  $B_D(u)$ in powers of $u$. Then the Borel 
integral is regular, no prescription is required and we obtain from (\ref{RBDm}), for $s<\Lambda^2$: 
\begin{equation}\label{RBDN} 
{\cal R}^{(N)}(s)= \frac{1}{\beta_0}\int\limits_{0}^{-\infty} {\rm e}^{-u/(\beta_0 a(s))} 
\,\frac{\sin\pi u}{\pi u}\, B_D^{(N)}(u) {\rm d} u\,=- \frac{1}{\beta_0}\int\limits_{0}^{\infty} {\rm e}^{-u/(\beta_0 
(-a(s))} \,\frac{\sin\pi u}{\pi u}\, B_D^{(N)}(-u) {\rm d} u\,,
\end{equation}
where a change of variable was performed in the last step. In the last integral the quantity $-a(s)$ is positive
 for $s<\Lambda^2$ (recall the definition (\ref{as1loop}) of the coupling), and we can easily perform the integration for 
 each term of $B_D^{(N)}(-u)$ from (\ref{BN}). A straightforward calculation  shows that, for all the terms except the first
 one, the infrared limit of the  integral (\ref{RBDN}) is zero, in agreement with the behavior of the  functions $A_n(s)$ of 
(\ref{An}), with $n>1$. For the  first term  $B_D(u)=1$ (which was responsible for the infrared freezing of the  truncated 
expansion in Section II) we apply the identity (41) of \cite{HoMa} which gives
 \begin{equation}\label{RBD1} 
 {\cal R}^{(0)}(s)=- \frac{1}{\beta_0}\int\limits_{0}^{\infty} {\rm e}^{-u/(\beta_0 (-a(s)))} 
\,\frac{\sin\pi u}{\pi u}\,  {\rm d} u\,= - \frac{1}{\pi\beta_0} \arctan [\pi\beta_0 (-a(s)]\,. 
\end{equation}
In the infrared 
limit $s\to 0$,  Eq. (\ref{as1loop})  implies$-a(s)\to 0$ through positive values, and from (\ref{RBD1})
 we obtain  ${\cal R}^{(0)}(0)=0$, which is not consistent with the infrared limit of the function $A_1(s)$ 
defined in (\ref{An}), and with the relation (\ref{IF}). Therefore, the prescription adopted in \cite{HoMa} for the
 Minkowskian quantity at $s<\Lambda^2$ fails to reproduce correctly the infrared freezing  (\ref{IF}) of the
 truncated expansion.
\section{Conclusions}
In a recent paper \cite{HoMa} it is claimed that by using analytic  continuation in the energy plane it
 is possible toprove the  infrared freezing of Minkowskian quantities beyondfinite-orders in perturbative QCD. 
 In the present work, we applied the technique of the inverse Mellin transform of the Borel function, developed in 
\cite{CaNe}, which givescompact expressions of the QCD amplitudes in the complex plane. By using the  analytic
 continuation of these expressions into the Landau region,  we calculated explicitly the spectral functions, as in the 
fixed order expansion. As in \cite{HoMa} we  adopted the Principal Value prescription, and considered as Euclidean quantity 
 the Adler function in massless QCD. Our result, expressed in Eq. (\ref{Rinfra}),  contradicts the conclusion  reached in 
\cite{HoMa}: the summation of higher orders in QCD leads to a divergent increase of the Minkowskian quantities in the infrared 
 limit, if these are calculated by analytic continuation from the Euclidean region. The divergent infrared behavior 
 arises explicitly from  the summation of the infinite terms and is related to the infrared renormalons. Of course, one 
expects that in full QCD this divergent behavior will  be compensated by a similar growth of remaining terms in the OPE, 
calculated with the same prescription. 

The difference between our results and those in \cite{HoMa}  is explained by  the 
fact that the authors of   \cite{HoMa}  do not apply consistently the  principle of analytic continuation, applied at
 finite orders. Lacking a  compact expression of the Adler function, as the one provided by the inverse  Mellin transform
 used by us, the authors make the analytic continuation of  the Borel representation itself, which is valid only outside 
the Landau region.  Therefore, the Borel representation (\ref{RBD}) of the Minkowskian quantity  considered in \cite{HoMa} 
 converges only for $s>\Lambda^2$, and is useless  in the infrared limit. To reach this point the authors change the
 definition  of the Borel integral. However, as we showed in Section V, this new prescription for the Borel summation
 of Minkowskian quantities below the  Landau point fails to reproduce the infrared freezing of the truncated  expansion
. 
\begin{acknowledgments}We thank Prof. Ji\v{r}\'{i} Ch\'{y}la  for many interesting and stimulating discussions on topics 
related to this work, and Dr. Chris J. Maxwell for a valuable correspondence. We are indebted to a referee for useful comments and interesting insights on some aspects of this work.   I.C. thanks Prof. Ji\v{r}\'{i} Ch\'{y}la  
and the Institute of Physics of the Czech Academy for hospitality. This work was  supported by the CERES Program of Romanian
 MEC under Contract  Nr. C3/125, by the Romanian Academy under Grant Nr. 20/2004, by the BARRANDE Project Nr. 2004-026-1, 
and by the Ministry of Education, Youth and Sports of  the Czech Republic, Project Nr. 1P04LA211. \end{acknowledgments}

\end{document}